\pdfoutput=1
\documentclass[conference]{IEEEtran}
\usepackage[utf8]{inputenc}
\usepackage{cite}
\usepackage{floatrow}
\usepackage{amsmath,amssymb,amsfonts,stfloats}
\usepackage{tabularx}
\usepackage{graphicx}
\usepackage{textcomp}
\usepackage{floatrow}
\usepackage{xcolor}
\usepackage{booktabs}

\usepackage{multicol}
\usepackage{acronym}
\usepackage{url}
\usepackage{mathtools}
\usepackage{algorithm}
\usepackage{algpseudocode} 
\usepackage{enumitem}
\usepackage[caption=false,font=footnotesize]{subfig}

\def\BibTeX{{\rm B\kern-.05em{\sc i\kern-.025em b}\kern-.08em
    T\kern-.1667em\lower.7ex\hbox{E}\kern-.125emX}}

\begin{document}

\title{On the Joint Estimation of Phase Noise and Time-Varying Channels for OFDM under High-Mobility Conditions}

\author{\IEEEauthorblockA{Francesco Linsalata\textit{$^1$} and Nassar Ksairi\textit{$^2$}}
\IEEEauthorblockA{\textit{$^1$}Dip. di Elettronica, Informazione e Bioingegneria, Politecnico di Milano, Milano
Italy (francesco.linsalata@polimi.it)}
\IEEEauthorblockA{\textit{$^2$}Mathematical and Algorithmic Sciences Lab, Huawei France R\&D, Paris, France (nassar.ksairi@huawei.com) }}

\maketitle

\begin{abstract}
The combination of the effects of Doppler frequency shifts (due to mobility) and phase noise (due to the imperfections of oscillators operating at a high carrier frequency) poses serious challenges to Orthogonal Frequency Division Multiplexing (OFDM) wireless transmissions in terms of channel estimation and phase noise tracking performance and the associated pilot overhead required for that estimation and tracking. In this paper, we use separate sets of Basis Expansion Model (BEM) coefficients for modelling the time variation over intervals of several OFDM symbols of the channel paths and the phase noise process. Based on this model, an efficient solution approximating the maximum-likelihood joint estimation of these BEM coefficients is derived and shown to outperform state-of-the-art phase noise compensation methods.
\end{abstract}

\section{Introduction}\label{Introduction}

Beyond the fifth generation (B5G) network's physical layer will have to cope with a high degree of heterogeneity in terms of services and deployment scenarios e.g., situations of high mobility and transmission using the millimeter wave (mmWave) frequency bands. 
In case of perfect synchronization between the transmitter (Tx) and the receiver (Rx), Orthogonal Frequency Division Multiplexing (OFDM) offers high-quality wireless communications due to its robustness to fading channels, adequate spectral efficiency, and easy integration with the multiple-input and multiple-output (MIMO) technology. However, when working at a high carrier frequency, a time-varying phase difference between the Tx and Rx local oscillators results in the so-called Phase Noise (PN).
The PN is responsible for a multiplicative part which is common to all the OFDM symbol subcarriers i.e., Common Phase Error (CPE), and an additive part at the subcarrier level that introduces Inter-Carrier Interference (ICI)~\cite{SamtaPNmmW, armada2001understanding, schenk2007distribution}. As proposed in 3GPP standards~\cite{3GPPNR,3GPPPTRS}, Phase Noise Tracking Reference Signal (PTRS) pilots allows for CPE estimation and mitigation~\cite{chungReview, sibel2018pilot, qi2018phase,zhang2021efficient}.
While CPE compensation results in performance improvement~\cite{sibel2018pilot}, ICI aware detection is needed to enable the use of high modulation and coding schemes (MCS)~\cite{KrishnaCons}.
Moreover, some of the upcoming B5G applications are expected to be deployed in mobility conditions, e.g. vehicular communications, where the fast channel variations also introduce ICI. An accurate channel state information (CSI) tracking these fast variations is needed under these conditions for ICI compensation and data recovery.
The combination of PN and mobility can easily degrade the quality of the link unless sufficient pilot overhead is used to enable PN and CSI estimation.
Under such conditions, a Basis Expansion Model (BEM) can typically well model the channel dynamics. Since the order of the BEM model needed for most practical channel representation accuracy values is smaller than the number of subcarriers, lower pilot overhead is needed hence guaranteeing a higher spectral efficiency~\cite{chen2009bem, idrees2010time, MatheckenICASSP, MatheckenPNGeometry}.
Since the Power Spectral Density (PSD) of PN decays rapidly beyond the loop bandwidth of the oscillator, the PN process can be sufficiently characterized by BEM consisting of a few lower-order spectral components, containing most of its energy. In~\cite{MATHECKEN2017,SyrjalaNonIter}, methods for PN estimation based on such a BEM representation of the PN variations within an OFDM symbol are proposed while assuming the channel to be time-invariant within the OFDM symbols and perfectly known. A different BEM was used in~\cite{sibel2018pilot} for the same purpose. the problem of time-invariant channel estimation and data detection in the presence of PN was addressed in \cite{SHRIVASTAV202155} using a similar per-OFDM-symbol PN BEM and the joint estimation of the channel, the PN BEM and the data symbols vector.
BEM-based estimation for time-varying channels in presence of PN was addressed in~\cite{ZangOFDMPNtimevar}. However, a single BEM was used to model channel variations within OFDM symbols under the cumulative effect of Doppler frequency spread and PN.

\subsection*{Paper Contributions}
The main contributions of this paper are here summarized.
\begin{itemize} [wide]
\item Most of the above-mentioned works address PN and CSI estimation as two distinct problems; and when they are addressed jointly, it is the cumulative effect of PN and Doppler that is modelled and estimated with the consequence of requiring large pilot overhead. By contrast, we propose a method to jointly tackle the time-varying channel and PN estimation that does not require such a large overhead.
\item A low-complexity algorithm for approximating the pilot-assisted Maximum Likelihood (ML) estimate for the above joint estimation problem is proposed. 
\item Using a BEM to model PN variations on the OFDM symbol level implicitly imposes the need for PN tracking pilots in every symbol. We instead solve the problem of estimating the coefficients of frame level\footnote{more precisely, per-sub-frame, using the 3GPP terminology} PN and channel BEMs based on pilots e.g., 3GPP Demodulation Reference Signal (DMRS)~\cite{3GPPNR} and PTRS~\cite{3GPPPTRS}, scattered throughout the frame.
\item Numerical results obtained using simulations with realistic channel models, 5G frame structure, and channel coding demonstrate the benefits of the proposed approach with respect to CPE-based technique in terms of block error rate (BLER) for mmWave communications with user mobility.
\end{itemize}

\section{System Model}
This section is dedicated to the baseband system model for an OFDM system experiencing PN and relative movement between the Tx and the Rx.
Consider a Single Input Single Output (SISO) OFDM transmission frame consisting of $M$ consecutive OFDM symbols, each consisting of $K$ subcarriers with a total system bandwidth of $B_{\max}=K{\Delta f}$ where ${\Delta f}=\frac{1}{T}$ is the subcarrier spacing and $T$ is the OFDM symbol duration. 
We designate the $K$-long vector of frequency domain samples of the $m$th OFDM symbol ($m\in\{0,\ldots,M-1\}$) as $\mathbf{s}_m$.
For the Rx to be able to estimate the communication channel matrix, DMRS and PTRS pilots are inserted by the Tx within the symbols $\left\{\mathbf{s}_m\right\}_{m=0,\ldots, M-1}$ following patterns defined by NR 5G specifications~\cite{3GPPNR, 3GPPPTRS}.
The time domain samples of the $m$th OFDM symbol $\mathbf{x}_m\in\mathbb{C}^{K\times1}$ for $m=0,\ldots, M-1$ is formed by applying the Inverse Discrete Fourier Transformation (IDFT) to the complex data symbols $s_k$ for ${k=0,\ldots,\ K-1}$, and by placing a copy of the last $K_{\rm cp}$ samples ($K_{\rm cp}<K$) in front of the symbol to form a cyclic prefix (CP) of length $K_{\rm cp}$. This can be written as 
$\mathbf{x}_m= \mathbf{A}_{\rm cp} \mathbf{F}^{\rm H} \mathbf{s}_m$
where $\mathbf{A}_{\rm cp}\in{\ \mathbb{R}}^{(K+K_{\rm cp})\times K}$ is the CP insertion matrix 
and $\mathbf{F}\in\mathbb{C}^{K\times K}$ is the DFT matrix.
The discrete-time propagation channel from the Tx to the Rx is modelled as the sum of $P$ paths with delays $\left\{l_p\right\}_{p=1\cdots P}$ (indexed in increasing order). The time domain (TD) received samples in presence of RX side PN are given by
\begin{equation} \label{eq:tdchannel}
\textstyle
    r_n=\sum_{p=1}^{P} p_{n}h_{p,n} x_{n-l_p}+w_n,
\end{equation}
where $p_n$ is the $n$-th PN sample, $h_{p,n}$ is the time-varying complex amplitude of the $p$-th path ($p\in\{1,\ldots,P\}$, $n\in{0,\ldots,N-1}$) with $N=M(K+K_{\rm cp})$, and $w_n$ is the $\sigma^2$-variance additive white Gaussian noise (AWGN). Here, we assumed that PN is only present on the Rx side. This is done only for the sake of readability. All the proposed estimation algorithms are valid in the case where PN is also present at the Tx side. Each PN sample is given as $ p_{n}=e^{j \vartheta_{n}}$, with $\vartheta_{n}$ defined by a first-order Autoregressive Model AR(1) with innovation variance ${\sigma^2}_{\rm PN}=4\pi B_{\rm 3dB,PN}{\Delta t}$, where $B_{\rm 3dB,PN}$ is the 3dB bandwidth of the PN~\cite{armada2001understanding}. In vector form, the PN is added by performing an element-wise multiplication between the received TD vector during the $m$-th OFDM symbol (${m=0\dots M-1}$) and the PN vector ${\mathbf{p}_m\in\mathbb{C}^{(K+K_{\rm cp})\times1}}$ associated with the same symbol and satisfying $[\mathbf{p}_m]_{n}=p_{k_m+n}$ ($n=0\dots K+K_{\rm cp}-1$, $k_m\triangleq m(K+K_{\rm{cp}})$) to get
\begin{align} \label{eq:rxvecttd}   \mathbf{r}_m=\text{diag}\left(\mathbf{p}_{m}\right)\mathbf{B}_{\rm cp}\mathbf{H}^{\rm TD}_m{\mathbf{A}_{\rm cp}\mathbf{F}}^{\rm H}{\mathbf{s}_m} +\ \mathbf{w}_{m},
\end{align}
where $\text{diag}\left(\mathbf{p}_{m}\right)$ is a diagonal matrix with $\mathbf{p}_{m}$ as its main diagonal, $\mathbf{w}_{m}\sim\mathcal{CN}\left(\mathbf{0},\sigma^2\mathbf{I}_{K}\right)$, $\mathbf{B}_{\rm cp}\in{\mathbb{R}}^{K\times (K+K_{\rm cp})}$ represents CP removal and $\mathbf{H}^{\rm TD}_m\in\mathbb{C}^{{(K}+K_{\rm cp})\times (K+K_{\rm cp})}$ is the TD channel matrix defined in \eqref{eq:matrix}. 
\begin{figure*}[ht!] 
\begin{minipage}[b]{.5\textwidth}
\footnotesize
\begin{align} \label{eq:matrix}
    \mathbf{H}^{\rm TD}_m=
    \left(\begin{matrix} 
     \vdots&0& \cdots &\cdots &\cdots\ &0
    \\ h_{1, k_m}&\vdots&\ddots &\cdots &\cdots &\vdots\
    \\ \vdots&h_{1, k_m+1}& \ddots\ &\ddots\ & \cdots&\vdots\
    \\  h_{P, k_m} & \ddots &\ddots\ &\ddots\ &\ddots &\vdots\
    \\ 0 &\ddots& \ddots& \ddots\ &\ddots &0\
    \\ \vdots &\cdots &h_{P, k_m+ K+K_{\rm{cp}}-l_p-1}&\cdots&h_{1, k_m+ K+K_{\rm{cp}}-l_1- 1} &\cdots \\\end{matrix}\right), \nonumber
\end{align}
\end{minipage}
\begin{minipage}[b]{.35\textwidth}
\begin{align} 
    \mathbf{T}=\left(\begin{matrix}\ 0&\cdots&0&0&1\\1&0&\cdots\ &\vdots&0\ \\0&1&\ddots&\ &\vdots\\\vdots&0&\ddots\ &\ 0&\ \\\ 0&\vdots&0&\ 1&0\\\end{matrix}\right)
\end{align}
\end{minipage}
\end{figure*}
Define $\mathbf{y}_m\triangleq\mathbf{F}\mathbf{r}_m$ as the vector of received frequency domain (FD) samples of the $m$-th OFDM symbol. After some manipulations applied to \eqref{eq:rxvecttd},
\begin{equation} \label{eq:rxsamplescomplete}
\textstyle
    \mathbf{y}_m =\mathbf{F} \sum_{p=1}^P\text{diag}\left(\mathbf{p}_{m}\right){\text{diag}\left(\mathbf{h}_{p,m}\right)}\mathbf{T}^{l_p}\mathbf{F}^{\rm H}\mathbf{s}_m+\mathbf{n}_{m} 
\end{equation}
where $\mathbf{T}\in{\ \mathbb{R}}^{K\times K}$ is the permutation matrix defined in \eqref{eq:matrix}.

\section{Basis Expansion Model for Time-Varying Channel and Phase Noise Estimation}

In BEM approaches \cite{chen2009bem}, $\mathbf{h}_p\stackrel{\rm def.}{=}\left[h_{p,n}\right]_{n=0\cdots N-1}$ consisting of the channel samples associated with the $p$-th term in \eqref{eq:tdchannel} is represented using ${\boldsymbol{\alpha}_{{\rm ch},p}\in\mathbb{C}^{(Q_{\rm ch}+1)\times1}}$ as
\begin{equation}
\label{eq:ch_bem}
    \mathbf{h}_p=\mathbf{B}_{\rm ch} \boldsymbol{\alpha}_{{\rm ch},p},
\end{equation}
where the columns of $\mathbf{B}_{\rm ch}\in\mathbb{C}^{{M(K}+K_{\rm cp})\times(Q_{\rm ch}+1)}$ are orthogonal basis sequences e.g.,
$\left[\mathbf{B}_{\rm ch}\right]_{n,q}=e^{j2\pi\left(q-\frac{Q_{\rm ch}}{2}\right)\frac{n}{N}}$ for exponential BEM. 
Thus, channel estimation boils down to estimating $P(Q_{\rm ch}+1)$ unknowns $\boldsymbol{\alpha}_{\rm ch}\triangleq\left[\boldsymbol{\alpha}_{{\rm ch},1}^{\rm T}\dots\boldsymbol{\alpha}_{{\rm ch},P}^{\rm T}\right]^{\rm T}$ based on {\it at least} $P(Q_{\rm ch}+1)$ transmitted pilot subcarriers.

\subsection{Channel and PN Estimation with a Single BEM} \label{sec:singelBEM}

In the presence of PN, the first approach is to use a single BEM to model the accumulative effect of mobility and PN on the channel.
For that sake, we rewrite \eqref{eq:rxsamplescomplete} as
\begin{align} \label{eq:rxsamplescomplete2}
    \mathbf{y}_m &=\mathbf{F}\textstyle\sum_{p=1}^P\mathbf{T}^{l_p}\ \text{diag}\left(\mathbf{h}_{p,m}\odot\mathbf{p}_{m}\right)\mathbf{F}^{\rm H}\mathbf{s}_m+\ \mathbf{n}_{m} \nonumber \\ 
    &=\mathbf{F}\ \textstyle\sum_{p=1}^P\mathbf{T}^{l_p} {\text{diag}(\mathbf{F}}^H \mathbf{s}_m)(\mathbf{p}_\mathbf{m} \odot \mathbf{h}_{p,m})+ \mathbf{n}_{m}
\end{align}
and we define ${\boldsymbol{\alpha}_{{\rm ch,pn},p}\in\mathbb{C}^{(Q_{\rm ch,pn}+1)\times1}}$ as the BEM representation of the point-wise product vector $\mathbf{p}_\mathbf{m} \odot \mathbf{h}_{p,m}$ associated with the basis matrix $\mathbf{B}_{\rm ch,pn}\in\mathbb{C}^{N\times(Q_{\rm ch,pn}+1)}$. 
Let $K^{\rm o}\leq MK$ designate the total number of DMRS and PTRS pilot subcarriers (including any necessary guard (null) subcarriers surrounding them) in the entire OFDM frame and define $\mathbf{y}^{\rm o}$ as the aggregate of the frame samples received at the positions of these subcarriers i.e., $\mathbf{y}^{\rm o}=\mathbf{A}^{\rm o}\left[\mathbf{y}_0 \ldots \mathbf{y}_{M-1}\right]^{\rm T}$ where matrix $\mathbf{A}^{\rm o}\in\mathbb{R}^{K^{\rm o}\times MK}$ has per row a single non-zero entry that is equal to one and which occupies the position of one of the pilots or guard subcarriers. Then
\begin{align} \label{eq:frame_bem}
\mathbf{y}^{\rm o}=\mathbf{S}\left(\mathbf{I}_{P}\otimes\mathbf{B}_{\rm ch,pn}\right)\boldsymbol{\alpha}_{\rm ch,pn}+ \mathbf{n}=\mathbf{z}+\mathbf{n},
\end{align}
where $\otimes$ stands for the Kronecker product, $\mathbf{n}$ is the additive noise vector and $\mathbf{S}\in{\mathbb{C}}^{K^{\rm o} \times MK}$ is the {\it sensing} matrix linking the unknown vector $\boldsymbol{\alpha}_{\rm ch,pn}^{\rm T}\triangleq\left[\boldsymbol{\alpha}_{{\rm ch,pn},1}^{\rm T} \dots \boldsymbol{\alpha}_{{\rm ch,pn},P}\right]^{\rm T}$ to the measurements vector. Referring to \eqref{eq:rxsamplescomplete2}, $\mathbf{S}$ can be written as a block row of the following sub-matrices ($m=0\dots M-1$)
\begin{align}
\label{eq:blockS}
\textstyle
    \left[\mathbf{S}\right]_{:,(m-1)K+1:(m-1)K+K}=\mathbf{A}^{\rm{o}}\mathbf{F}\ \mathbf{T}^{l_p} {\text{diag}(\mathbf{F}}^H \mathbf{s}_{m}^{\rm o}),
\end{align}
where $\mathbf{s}_{m}^{\rm o}$ equals $\mathbf{s}_{m}$ at pilot subcarriers and zero elsewhere.
Due to the i.i.d. Gaussian nature of $\mathbf{n}$, the log-likelihood function for estimating $\boldsymbol{\alpha}_{\rm ch,pn}$ from $\mathbf{y}^{\rm o}$ is $\mathcal{L} (\boldsymbol{\alpha}_{\rm ch,pn})=\left\|\mathbf{y}^{\rm o} -\mathbf{z}\right\|^2$ and the maximum likelihood estimate ${\widehat{\boldsymbol{\alpha}}}_{\rm ch,pn}\triangleq \text{argmin}_{\boldsymbol{\alpha}_{\rm ch,pn}} \mathcal{L} (\boldsymbol{\alpha}_{\rm ch,pn})$ is ($^{\dagger}$: Moore-Penrose inverse)
\begin{align} \label{eq:MLframe_singleBEM}
{\widehat{\boldsymbol{\alpha}}_{\rm ch,pn}}=\left(\mathbf{S}\left(\mathbf{I}_{P}\otimes\mathbf{B}_{\rm ch,pn}\right)\right)^{\dagger}\mathbf{y}^{\rm o}.
\end{align}
The advantage of the single-BEM formulation is that the ML estimate as given by \eqref{eq:MLframe_singleBEM} is linear in $\mathbf{y}^{\rm o}$, thus guaranteeing low complexity.
However, the pilot overhead becomes prohibitive, as validated in the simulations section, as soon as the $B_{\rm 3dB,PN}$ is large enough. Indeed, the entries of the vector $\mathbf{h}_p\odot\mathbf{p}$ can be seen as the samples of a time-varying channel with increased Doppler spread. Modelling and estimating such a channel requires a BEM with a higher order ($Q_{\rm ch,pn}>Q_{\rm ch}$) and, consequently, a larger pilot overhead.
\begin{figure*}[hb!] 
\begin{align} \label{eq:rxsamplescomplete3}
       \mathbf{y}_m  \triangleq \underbrace{\mathbf{F}\textstyle \sum_{p=1}^P \text{diag}\left(\mathbf{A}_{\rm{m}}\mathbf{B}_{\rm pn} \boldsymbol{\alpha}_{{\rm pn},p}\right) {\text{diag}\left(\mathbf{A}_{\rm{m}}\mathbf{B}_{\rm{ch}} \boldsymbol{\alpha}_{{\rm{ch}},p}\right)}\mathbf{T}^{l_p} \mathbf{F}^{\rm H}\mathbf{s}_m}_{\mathbf{z}_m\left(\mathbf{s}_m,\boldsymbol{\alpha}_{{\rm pn}},\boldsymbol{\alpha}_{{\rm ch}}\right)} +\mathbf{n}_{m} 
\end{align} 
\end{figure*}

\subsection{Channel and PN Estimation with Separate BEMs}

We now model the PN vector $\mathbf{p}$ using a separate BEM as
\begin{equation}
\label{eq:pn_bem}
    \mathbf{p}=\mathbf{B}_{\rm pn}\boldsymbol{\alpha}_{\rm pn}
\end{equation}
with $\boldsymbol{\alpha}_{\rm pn}\in\mathbb{C}^{
Q_{\rm pn}\times1}$. Equation \eqref{eq:rxsamplescomplete} can be rewritten as \eqref{eq:rxsamplescomplete3} where $\mathbf{A}_{m}$ is a $\mathbb{R}^{K\times N)}$ selection matrix that is defined according to the $m$th OFDM symbol.
Since \eqref{eq:rxsamplescomplete3} is no longer linear, an iterative approach is here investigated to estimate both the channel taps $\mathbf{h}_p$ and the PN samples $\mathbf{p}$.
In fact, the log-likelihood function $\mathcal{L}\left(\boldsymbol{\alpha}_{\rm pn}, \boldsymbol{\alpha}_{\rm ch}\right)= \left\|\mathbf{y}^{\rm o} -\mathbf{z}(\boldsymbol{\alpha}_{\rm pn}, \boldsymbol{\alpha}_{\rm ch})\right\|^2,$ with $\mathbf{z}(\boldsymbol{\alpha}, \boldsymbol{\beta})\triangleq \mathbf{A}^{\rm o}\times$ $[\mathbf{z}_0^{\rm T}(\mathbf{s}_0^{\rm o},\boldsymbol{\alpha},\boldsymbol{\beta})$ $\dots$ $\mathbf{z}_{M-1}^{\rm T}(\mathbf{s}_{M-1}^{\rm o},\boldsymbol{\alpha},\boldsymbol{\beta})]^{\rm T}$
is quadratic in both the PN $\boldsymbol{\alpha}_{\rm pn}$ and the channel $\boldsymbol{\alpha}_{\rm ch}$.

One possible approach to search for $\textrm{argmin}\mathcal{L}\left(\boldsymbol{\alpha}_{\rm pn}, \boldsymbol{\alpha}_{\rm ch}\right)$ is {\it alternating optimization} \cite{BEZDEK2003}: during iteration $t\geq1$ we derive $\boldsymbol{\alpha}_{\rm pn}^{(t)}$ that minimizes $\mathcal{L}\left(\boldsymbol{\alpha}_{\rm pn}, \widehat{\boldsymbol{\alpha}}_{\rm ch}^{(t-1)}\right)$ then $\widehat{\boldsymbol{\alpha}}_{\rm ch}^{(t)}$ that minimizes
$\mathcal{L}\left(\widehat{\boldsymbol{\alpha}}_{\rm pn}^{(t)}, \boldsymbol{\alpha}_{\rm ch}\right)$ until a stopping condition is satisfied. Note that ${\widehat{\boldsymbol{\alpha}}_{\rm pn}}^{(t)}$ can be found by using \eqref{eq:MLframe_singleBEM} with $\mathbf{B}_{\rm pn}$ replacing $\mathbf{B}_{\rm ch,pn}$ and with $\mathbf{S}$ defined using \eqref{eq:blockS} where its right-hand side (rhs) is $\mathbf{A}^{\rm{o}}\mathbf{F}\sum_{p=1}^P\text{diag}\left(\mathbf{A}_{\rm{m}}\mathbf{B}_{\rm ch} \boldsymbol{\alpha}_{{\rm ch},p}^{(t-1)}\right)\mathbf{T}^{l_p} \text{diag}\left(\mathbf{F}^{\rm H}\mathbf{s}_m^{\rm o}\right)$.
Then, $\widehat{\boldsymbol{\alpha}}_{\rm ch}^{(t)}$ can similarly be computed using \eqref{eq:MLframe_singleBEM} with $\mathbf{B}_{\rm ch}$ replacing $\mathbf{B}_{\rm ch,pn}$ and with $\mathbf{S}$ defined using \eqref{eq:blockS} with its rhs replaced by $\mathbf{A}^{\rm{o}}\mathbf{F}\sum_{p=1}^P\mathbf{T}^{l_p} {\text{diag}(\mathbf{F}}^H \mathbf{s}_{m}^{\rm o}) \text{diag}\left(\mathbf{A}_{\rm{m}}\mathbf{B}_{\rm pn} \boldsymbol{\alpha}_{{\rm pn},p}^{(t)}\right)$.
The pseudo-code of this procedure is reported in Algorithm~\ref{alg:loop}. The conditions of \cite[Theorems 2 and 3]{BEZDEK2003} are met\footnote{A technicality \cite{BEZDEK2003} requires the feasible solutions set to be compact which, strictly speaking, is not the case of $\mathbb{C}^{Q_{\rm ch}+Q_{\rm pn}+2}$. This can be alleviated by constraining $\left(\boldsymbol{\alpha}_{{\rm ch}},\boldsymbol{\alpha}_{{\rm pn}}\right)$ to lie in a ``large-enough'' compact subset.}, then Algorithm \ref{alg:loop} provides a sequence of non-increasing log-likelihood values and an associated sequence of BEM coefficients that converge at least to a local solution to $\textrm{argmin}\mathcal{L}\left(\boldsymbol{\alpha}_{\rm pn}, \boldsymbol{\alpha}_{\rm ch}\right)$. 
Once the stopping condition is met, the estimated PN samples are used to compensate the PN in the received signal vector, while the frequency domain channel matrix (computed from the estimated $\mathbf{h}_p$) is used for linear minimum mean square error detection of $\left\{\mathbf{s}_m\right\}_{m=0\cdots M-1}$.

\begin{algorithm}[!hpbt] 
\caption{Channel and Phase Noise Estimation}
\label{alg:loop}
\begin{algorithmic}
\Statex \textbf{Input:} {$Q_{\rm ch}$, $Q_{\rm pn}$, $t_{\max}$, $\epsilon$}, \
 \textbf{Output:} {$\{\mathbf{h}_p^{(t)}\}_{p=1\dots P}$ and $\mathbf{p}^{(t)}$} 
\Statex \textbf{init} $[\mathbf{h}_p^{(0)}]_n\leftarrow 1 \,\, \forall p=1\cdots P, n=0\cdots N-1$
\Statex \textbf{for} $t\ =\ 1,\ 2,\ \ldots$ \textbf{do}
 \Statex  $\widehat{\boldsymbol{\alpha}}_{\rm pn}^{(t)}\leftarrow\mathrm{argmin}\ \mathcal{L}(\boldsymbol{\alpha}_{\rm pn}, \widehat{\boldsymbol{\alpha}}_{\rm ch}^{(t-1)})$, $\mathbf{p}^{(t)}\leftarrow\mathbf{p}(\widehat{\boldsymbol{\alpha}}_{\rm pn}^{(t)})$ \eqref{eq:pn_bem}
\Statex $\widehat{\boldsymbol{\alpha}}_{\rm ch}^{(t)}\leftarrow\mathrm{argmin} \ \mathcal{L}(\widehat{\boldsymbol{\alpha}}_{\rm pn}^{(t)}, \boldsymbol{\alpha}_{\rm ch})$, $\mathbf{h}_p^{(t)}\leftarrow\mathbf{h}_p(\widehat{\boldsymbol{\alpha}}_{\rm ch}^{(t)})$ \eqref{eq:ch_bem}
 \Statex \textbf{end} {when $t=t_{\max}$ or $\|\widehat{\boldsymbol{\alpha}}_{\rm ch}^{(t)}-\widehat{\boldsymbol{\alpha}}_{\rm ch}^{(t-1)}\|^{2}<\epsilon$}
\end{algorithmic}
\end{algorithm}
\begin{table}[htpb]
\begin{small}
\centering
\caption{{Simulation variables}}\label{matlabparameters}
\begin{tabular}{ | c | c | p{42mm} |}
	\hline 
	\textbf{Parameter}  & \textbf{Value}  \\ \hline
(Modulation size, LDPC coding rate) & (64,1/2) \\ \hline
Channel model & 3GPPTDLC100 \\ \hline
$(K,K_{\rm cp},M)$ & (144,36,28) \\ \hline
$B_{\max}$, Transmission bandwidth & 8.640 MHz \\ \hline
$f_{\rm c}$, Carrier frequency & 30 GHz \\  \hline
$B_{\rm 3dB,\ PN}$, 3dB bandwidth of the PN & 450 Hz \\ \hline
BEM type & DPSS\footnote{Discrete Prolate Spheroidal Sequences (DPSS) as in \cite{dpss} }\\ \hline
	\end{tabular}
	\label{tab:sim_setting}
 \end{small}
\end{table}
\begin{figure}[!t] 
\centering
\subfloat[CPE \label{fig:PN_cpe}]{\includegraphics[width=\columnwidth]{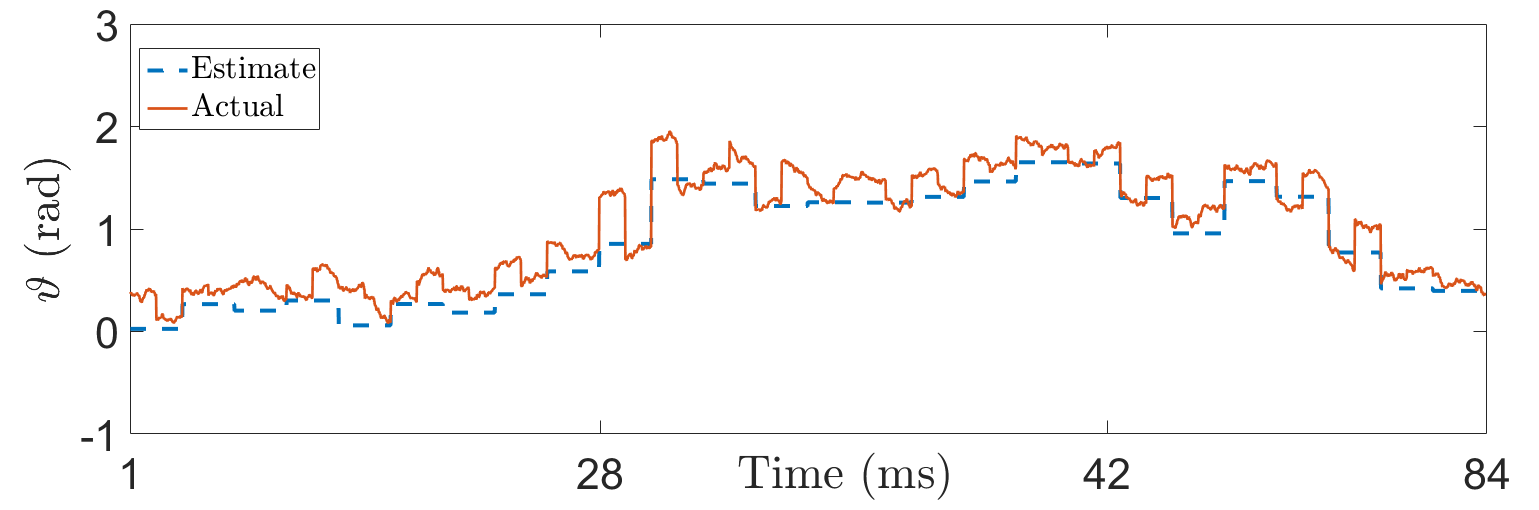}}

\subfloat[BEM \label{fig:PN_bem}]{ \includegraphics[width=\columnwidth ]{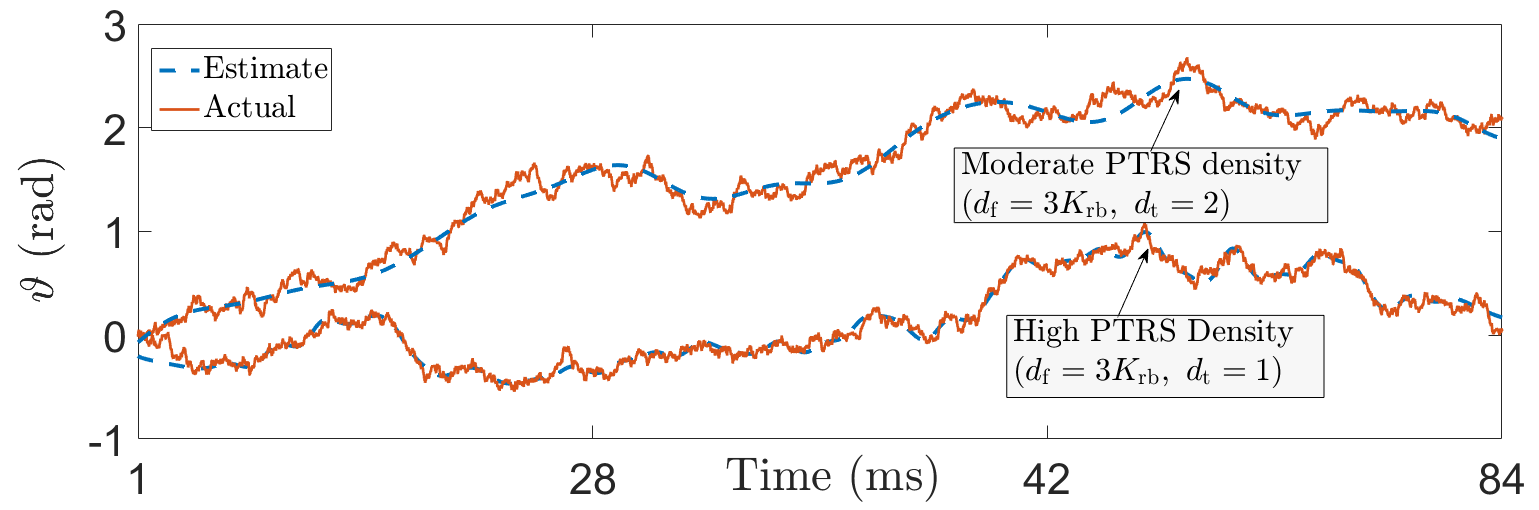}}
\caption{Actual and estimated PN samples (a) CPE, (b) BEM }
\label{fig:PN_samples}
\end{figure}
\begin{figure}[!t] 
\subfloat[][Rx speed $v_u= 3 \ km/h$ \label{fig:64_low_joint}]
{\includegraphics[width=\columnwidth]{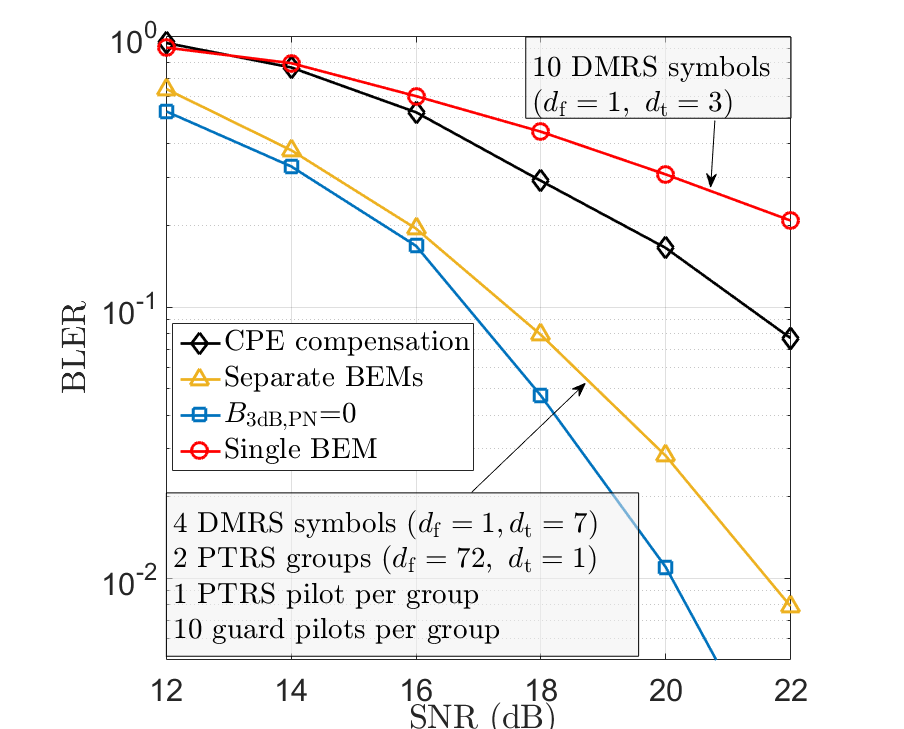}}

\subfloat[][Rx speed $v_u= 30 \ km/h$ \label{fig:64_high_joint}]{ \includegraphics[width=\columnwidth]{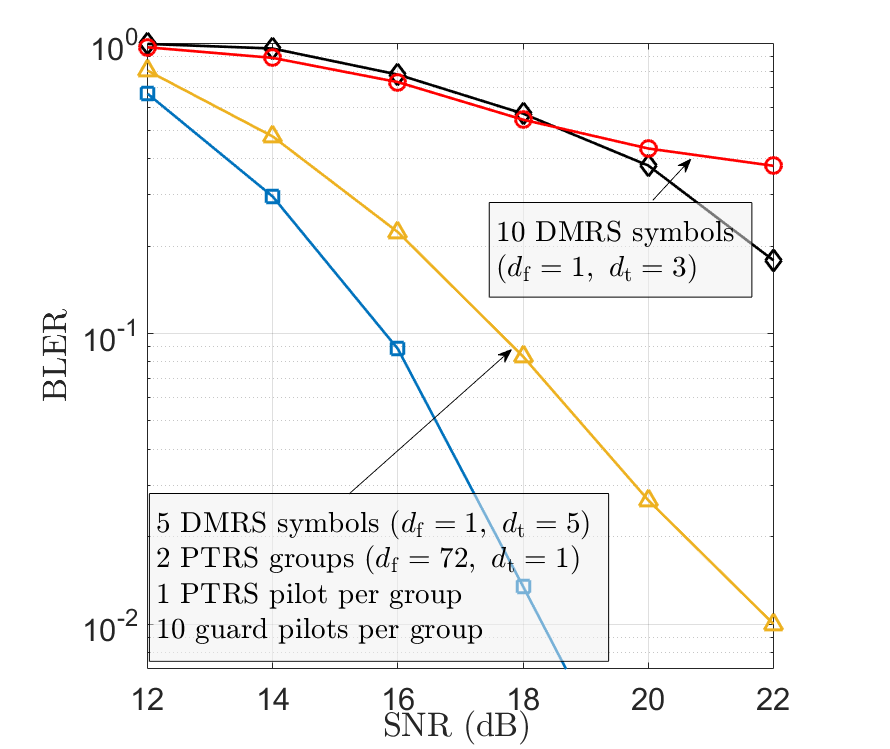}}
\caption{ 64 QAM BLER versus SNR in a) low mobility and b) moderate mobility with iterative CSI and PN estimation}
\label{fig:joint64}
\end{figure}
\section{Numerical Results}

This section provides numerical evaluations of the two approaches described above and compares them to the method of CPE estimation, widely studied in the literature~\cite{armada2001understanding} and consisting in approximating the PN by the CPE i.e., the common rotation of all the subcarriers within an OFDM symbol, it induces. This implies that ICI due to PN is ignored. CPE-based PN estimation methods are equivalent to a BEM-based method with $Q_{\rm pn}=0$.
The main simulation parameters are reported in Table \ref{tab:sim_setting}.
The performance metrics used for evaluation are BLER versus the Signal-to-Noise Ratio (SNR) and the MSE between the estimated and real PN samples. 
The benchmark is given by $B_{\rm 3dB, PN}=0$.
Let $K_{\rm rb}$ be the number of subcarriers in a resource block i.e., 12 in LTE and 5G, and denote by $d_{\rm f}$ and $d_{\rm t}$ the frequency domain (in subcarriers) and the time domain (in OFDM symbols) density of PTRS pilots, respectively. Figure~\ref{fig:PN_samples} (obtained using one iteration of Algorithm \ref{alg:loop} initialized with the actual channel) compares the CPE-based and BEM-based PN estimates versus the actual samples. The CPE-based estimates are constant within OFDM symbols, and unable to track shorter-term fluctuations. BEM-based PN estimates, both in case of high ($d_{\rm f}=3 K_{\rm rb},\ d_{\rm t}=1$) and moderate ($d_{\rm f}=3K_{\rm rb},\  d_{\rm t}=2$) PTRS density, are able to track the PN samples, guaranteeing a lower MSE. 
The performance of Algorithm \ref{alg:loop} is compared Fig.~\ref{fig:joint64} with single-BEM PN-channel estimation (Sec.~\ref{sec:singelBEM}) and the CPE-based method for both low and high mobility. The proposed per-frame separate-BEMs approach yields the best BLER performance. Interestingly, the excessive requirement of the single-BEM approach in terms of pilot overhead makes its performance worse than the CPE-based method under realistic pilot patterns (10 DMRS symbols in every frame).

\section{Concluding remarks}

In this paper, we investigated the problem of the joint channel and PN estimation for an OFDM transmission with both Rx mobility and high carrier frequency. We proposed an approach where the first set of BEM coefficients models the time-variation over a frame of OFDM symbols of the channel paths, while the second one models the PN realization over the same interval. We provided an efficient approximate solution to the problem of ML joint estimation of these BEM coefficients. Numerical results showed the advantage of the proposed solution over current PN compensation methods.

\newpage
\bibliographystyle{IEEEtran}
\bibliography{Bibliography}

\end{document}